\documentclass[aps,twocolumn,amssymb,showpacs,amsmath,nfootinbib,preprintnumbers]{revtex4}

\usepackage{graphicx,bm}

\def\bra#1{\langle #1|}
\def\ket#1{|#1\rangle}
\def\bracket#1{\langle #1 \rangle}

\def\sub#1{_{\mathrm{#1}}}

\def\3He{\mbox{$^3$He}}
\def\4He{\mbox{$^4$He}}

\begin{document}
\vspace{5cm}

\title{Spacetime analogue of Bose-Einstein condensates: Bogoliubov-de Gennes formulation}


\author{Yasunari Kurita$^{1}$, Michikazu Kobayashi$^{2}$, Takao Morinari$^3$, Makoto Tsubota$^4$, Hideki Ishihara$^4$}

\affiliation{$^1$
Department of Physics, Kwansei Gakuin University, Sanda, Hyogo 669-1337, Japan}

\affiliation{$^2$
Department of Physics, The University of Tokyo, Tokyo 113-0033, Japan}

\affiliation{$^3$
Yukawa Institute for Theoretical Physics, Kyoto University, Kyoto 606-8502, Japan}

\affiliation{$^4$
Department of Mathematics and Physics, Osaka City University, Osaka 558-8585, Japan}

\vspace{2cm}

\begin{abstract}
We construct quantum field theory in an analogue curved spacetime 
in Bose-Einstein condensates based on the Bogoliubov-de Gennes equations, 
by exactly relating quantum particles in curved spacetime with 
Bogoliubov quasiparticle excitations in Bose-Einstein condensates.
Here, we derive a simple formula relating the two, which can be used to 
calculate the particle creation spectrum 
by solving the time-dependent Bogoliubov-de Gennes equations.
Using our formulation, we numerically investigate particle creation 
in an analogue expanding Universe which can be expressed as 
Bogoliubov quasiparticles in an expanding Bose-Einstein condensate. 
We obtain its spectrum, which follows the thermal Maxwell-Boltzmann distribution, 
the temperature of which is experimentally attainable.
Our derivation of the analogy is useful for general Bose-Einstein condensates and
not limited to homogeneous ones, and our simulation is the first
example of particle creations by solving the Bogoliubov-de Gennes
equation in an inhomogeneous condensate.
\end{abstract}

\preprint{OCU-PHYS-305}
\preprint{AP-GR-62}
\preprint{YITP-08-77}

\pacs{03.75.Kk, 03.75.Hh, 05.30.Jp, 04.62.+v}

\maketitle
\section{Introduction}

Since the establishment of the theory of general relativity by A. Einstein, 
the study of classical gravity has been greatly extended. 
It provides some fascinating research subjects, especially cosmology and black holes.
These two topics are powerful motivations for studying physics in curved spacetime.

The quantum effects of matter fields in curved spacetime is also important 
for studying cosmology and black hole physics.
In standard modern cosmology, our Universe began with an accelerated expansion, 
known as inflation. 
The quantum effect of matter fields in the inflationary universe is believed to 
give rise to cosmological structures such as galaxies. 
However, this is merely theoretical scenario lacking any experimental evidence. 
Another example relates to the very mysterious objects known as black holes.
According to S. W. Hawking's prediction~\cite{hawking}, 
black holes are not completely black but rather they emit black body radiation 
through particle creation in the curved spacetime.
It is very difficult to confirm this surprising phenomenon experimentally due to 
the extreme low intensity of the emitted radiation.

With such a background, new attempts to discover analogous phenomena 
in fluid systems have appeared.
In a pioneering study, Unruh found 
that Hawking radiation is not restricted to gravity fields but that 
it can also be observed in a perfect fluid with a sonic horizon~\cite{unruh81}.
The basic concept involves identifying a sound wave propagating 
in a flowing fluid with a scalar field in curved spacetime.
In this identification, fluid flow corresponds to 
curved spacetime and provides the spacetime metric.
The sound wave can then be considered as a field in curved spacetime.
This analogy has been applied not only to Hawking radiation 
but also to other field theoretical phenomena in curved spacetime, 
including cosmological particle creation~\cite{novello2002,Barcelo:2005fc}.

One of the merits of considering such an analogy lies in possibility of 
observing quantum effects in a laboratory. 
For the purpose of performing a precise investigation of quantum effects, 
we should consider quantum fluids.
Bose-Einstein condensates (BECs) in trapped cold atoms represent 
one of the best systems in this context~\cite{Anderson-95,Ketterle95,Pethick}, 
for the following three reasons: 
(i) a BEC system is quite dilute and weakly interacting, 
so that its microscopic theory can be easily studied, 
(ii) many physical parameters of BECs are experimentally controllable, 
and (iii) various quantum phenomena of BECs can be directly observed, 
in stark contrast with other quantum fluids, such as superfluid helium systems.
Therefore, BECs are one of the most promising quantum fluids for an analogous system.

In this spacetime analogy, spacetime corresponds to the condensate wave function 
described by the Gross-Pitaevskii (GP) equation~\cite{GP,Gross}. 
Furthermore, quanta of the scalar field in spacetime correspond 
to Bogoliubov quasiparticles, which are low energy excitations on condensates 
and can be described by the Bogoliubov-de Gennes (BdG) equations.
Bogoliubov quasiparticles not only exist in BECs at finite temperatures 
but they are also spontaneously created through the motion of BECs, 
which is one of the important problems associated with defining a BEC 
and its superfluidity in nonequilibrium Bose systems.

One of the main subjects of the analogy in BEC is to verify the Hawking radiation~\cite{Garay-PRL,
Garay:2000jj,Barcelo:2001ca,
Leonhardt02-1,Leonhardt02-2,Giovanazzi04,Giovanazzi:2004zv,Wuester07,Kurita:2007bv,
Balbinot07,Carusotto08,Jain-PRA07,Wuester0805}. 
Not only the Hawking effect, other aspects of black hole physics as 
quasinormal frequencies or superradiance were also investigated~\cite{Nakano:2004ha,
Basak:2005fv,Federici:2005ty,Barcelo:2006yi,Takeuchi07,Barcelo:2007ru}.
Another stream is on cosmological models representing 
homogeneous expanding universe~\cite{Barcero02,Barcero03,Lidsy03,Jain07,Weinfurtner:2008if,Uhlmann0810}
or inflationary universe~\cite{Uhlmann0810,Fedichev03-1,
Fedichev03-2,Fedichev03-3,Fischer04,Weinfurtner04,Uhlmann:2005hf}.
The analogy in BEC was firstly formulated by Garay {\it et al.}~\cite{Garay-PRL}
and explored by Barcel\'{o} {\it et al.}~\cite{Barcelo:2000tg}. 
In their formulation, the analogue geometry is read from a second-order 
differential equation for the perturbed phase of the condensate wave 
function which obeys the GP equation.
In order to connect quantum physics in curved spacetime to the behavior of 
a realistic quantum fluid, Leonhardt {\it et al.}~\cite{Leonhardt02-1} 
investigated the Hawking effect within the 
Bogoliubov theory of the elementary excitations in BEC.
More detailed correspondence between the quanta in analogue spacetime
and the Bogoliubov quasiparticles
was discussed by Jain {\it et al.}~\cite{Jain07}, 
giving an analytical expression for particle creation spectrum in terms 
of the Bogoliubov mode functions in the case of homogeneous BEC.

In this paper, we formulate the spacetime analogy in terms of the BdG 
equations as an extension of \cite{Leonhardt02-1}, by exactly relating 
Bogoliubov quasiparticles with quantum particles in curved spacetime. 
It is shown that the number operator of quanta in analogue spacetime 
is different from that of Bogoliubov quasiparticles, unless the corresponding 
field is normalized correctly. 
With the correct normalization, the number of 
quanta in analogue spacetime can be exactly related with that of Bogoliubov quasiparticles.
Furthermore, it is shown that the orthonormal relation between
the normalized mode functions exactly corresponds to that of 
BdG wave functions.
We also explicitly define Bogoliubov quasiparticle creation as a phenomenon 
caused by a change in the complete set of BdG wave functions for the motion of a BEC, 
and we define analogue particle creation in terms of Bogoliubov quasiparticle creation.
Our formulation based on the BdG equations is generally applicable and 
gives a simple formula for particle creation spectrum in terms of the Bogoliubov mode functions. 
In the derivation of the formula, we consider the correctly normalized 
field so that the formula is relevant quantitatively. 
Our derivation of the formula is different from that given in~\cite{Jain07}, 
and consequently the formula is generalized to be applicable in general cases.
Furthermore, our formulation is independent of whether condensates 
are homogeneous or inhomogeneous, 
and the obtained formula is valid even for inhomogeneous condensates.

We then perform a numerical simulation for the simplest quantum effect 
in analogue curved spacetime: particle creation by the analogue expanding Universe 
as an expanding BEC.
The spectrum obtained is consistent with the thermal Maxwell-Boltzmann distribution, 
the temperature of which has already been achieved in actual experiments. 
In this simulation, 
we set the initial state for the excitation field to be vacuum just 
after the condensate starts to expand, and 
there is no external manipulation after the initial time
and the particle creation occurs spontaneously, 
which should be contrasted to 
the previous works~\cite{Barcero02,Barcero03,Lidsy03,Jain07,Weinfurtner:2008if}.
In this sense, the system is isolated. 
The particle creation simulated here is not a result of any perturbation
but a field theoretical phenomenon caused by a difference in definition 
of quasiparticles.

This paper is organized as follows.
In section \ref{sec:analogue}, we derive an analogue spacetime metric from the BdG equations.
In section \ref{sec:QFT-BdG}, we construct quantum field theory on 
the analogue spacetime based on BdG theory.
In section \ref{sec:particle-creation}, we define Bogoliubov quasiparticle creation 
and give a formula for analogue particle creation.
In section \ref{sec:numerical}, we show the results of a numerical simulation giving 
particle creation in an expanding and shrinking BEC in a trapping potential.
The conclusions are given in section \ref{sec:summary}.

\section{Analogue spacetime in BEC }
\label{sec:analogue}

We start with the Hamiltonian describing bosons trapped by an external 
confining potential $V\sub{ext}({\bf x})$ given by
\begin{align}
H = \int d^3x \: \psi^{\dagger}({\bf x}) \bigg[ & K + V\sub{ext}({\bf x}) 
 - \mu \nonumber \\
& + \frac{1}{2} U_0 \psi^{\dagger}({\bf x}) \psi({\bf x}) \bigg] 
\psi({\bf x}).
\end{align}
Here $\psi^{\dagger}({\bf x})$ and $\psi({\bf x})$ are boson field operators 
in the second quantized form, $K = - \hbar^2 \nabla^2 / 2m$ is the kinetic energy operator, 
$U_0$ is the particle interaction strength 
and $\mu$ is the chemical potential \cite{FW71,Fetter1972}.
In the Heisenberg picture, the field equation is given by
\begin{align}
i \hbar \partial_t \psi({\bf x}, t) & = [ K + V\sub{ext}({\bf x}) - \mu \nonumber \\
& \quad + U_0 \psi^{\dagger}({\bf x}, t) \psi({\bf x}, t)] \psi({\bf x}, t).
\end{align}
The density operator and the current density operator are respectively defined by
\begin{align}
\hat{n}({\bf x}, t) & = \psi^{\dagger}({\bf x}, t) \psi({\bf x}, t), \\
\hat{\bf j} ({\bf x}, t) & = \frac{\hbar}{2mi} [  \psi^{\dagger}({\bf x}, t) 
\nabla \psi({\bf x}, t) \nonumber \\
& \qquad \qquad - \{ \nabla \psi^{\dagger} ({\bf x}, t) \} \psi ({\bf x}, t) ].
\end{align}

In a BEC, the field $\psi$ consists of two components, 
$\psi({\bf x}, t) = \Psi({\bf x}, t) + \phi({\bf x}, t)$, 
where $\Psi({\bf x}, t) \equiv \langle \psi({\bf x}, t) \rangle$ 
describes the condensate \cite{PPO} with $\bracket{O}$ denoting 
a grand canonical ensemble average of $O$.
The Bogoliubov field $\phi({\bf x}, t)$ describes fluctuations about the condensate.
The condensate wave function $\Psi({\bf x}, t)$ obeys the following GP equation \cite{GP,Gross},
\begin{align}
i \hbar \partial_t \Psi({\bf x}, t) & = 
\left[ K + V\sub{ext}({\bf x}) - \mu \right] \Psi({\bf x}, t) \nonumber \\
& \quad + U_0 |\Psi({\bf x}, t)|^2 \Psi({\bf x}, t).
\end{align}

The condensate wave function $\Psi({\bf x}, t)$ is rewritten in terms of the amplitude, 
$|\Psi({\bf x}, t)| = \sqrt{n_0({\bf x}, t)}$, and the phase, $S({\bf x}, t)$, as follows,
\begin{align}
\Psi({\bf x}, t) = \sqrt{n_0({\bf x}, t)} e^{iS({\bf x}, t)}.
\end{align} 
Hereafter, the spatial and time dependence of fields, represented by $({\bf x}, t)$, 
is implicit for conciseness of notation.
The condensate contributions to the density and current density are
\begin{align}
n_0 = |\Psi |^2, \quad {\bf j}_0 = n_0 \frac{\hbar}{m} \nabla S = n_0  {\bf v}_0,
\end{align}
where ${\bf v}_0 := \hbar \nabla S / m$ is the velocity of the condensate.
With these quantities, the GP equation can be rewritten as two real equations
\begin{align}
\hbar \partial_t S + n_0^{-1/2} K & n_0^{1/2} + V\sub{ext} - \mu + U_0n_0
 + \frac{m}{2} v_0^2 = 0, \label{eq:GP-S} \\
 & \partial_t n_0 + \nabla \cdot (n_0 {\bf v}_0 ) = 0. \label{eq:conserv-0}
\end{align}
The first equation is Bernoulli's equation for the condensate current 
and the second equation represents conservation of the current.

Density fluctuations and current density fluctuations are described by
\begin{align}
\hat{\rho}^{\prime} & = \Psi^\ast \phi + \phi^{\dagger} \Psi 
= \sqrt{n_0} (e^{-iS} \phi + e^{iS} \phi^{\dagger} ), \\
\hat{\bf j}^{\prime} & = \frac{\hbar}{2m i}
 [ \Psi^\ast \nabla \phi + \phi^{\dagger} \nabla \Psi - (\nabla \Psi^\ast) \phi 
- (\nabla \phi^{\dagger} ) \Psi ].
\end{align}
Here, we take the linearized approximation, that is, 
we neglect higher order fluctuations, like $\phi^{\dagger} \phi$.
The fluctuation fields $\phi, \phi^{\dagger}$ obey the BdG equations.
These fields can be U(1) gauge transformed as
\begin{align}
\tilde{\phi} := e^{-iS} \phi, \quad \tilde{\phi}^{\dagger} := e^{iS} \phi^{\dagger}.
\end{align}
The field equations are
\begin{align}
i \hbar \partial_t \tilde{\phi} & = (\tilde{L} + 2 U_0 n_0 
+ \hbar \partial_t S)\tilde{\phi} + U_0 n_0 \tilde{\phi}^{\dagger}, \label{eq:B-dG-tilde-0} \\
- i \hbar \partial_t \tilde{\phi}^{\dagger} & = 
(\tilde{L}^{\ast} + 2 U_0 n_0 + \hbar \partial_t S) \tilde{\phi}^{\dagger} 
+ U_0 n_0 \tilde{\phi}, \label{eq:B-dG-tilde}
\end{align}
where $\tilde{L}$ is defined as
\begin{align}
\tilde{L} & := \tilde{K} + V_{ext} -\mu, \quad \tilde{K} := e^{-iS} K e^{iS}.
\end{align}
Performing a short calculation, one obtains
\begin{align}
\tilde{K}  = K -i \hbar {\bf v}_0 \cdot \nabla - 
\frac{1}{2} i \hbar ( \nabla \cdot {\bf v}_0 ) + \frac{1}{2}m v_0^2.
\end{align}
With these fields, $\hat{\rho}^{\prime}$ and $\hat{\bf j}^{\prime}$ take the simple forms
\begin{align}
\hat{\rho}^{\prime} = \sqrt{n_0} (\tilde{\phi} 
+ \tilde{\phi}^{\dagger}), \quad \hat{\bf j}^{\prime} = 
\hat{\rho}^{\prime}{\bf v}_0 + n_0 \hat{\bf v}^{\prime},
\end{align}
where we have introduced $\hat{\bf v}^{\prime} = \nabla \hat{\Phi}^{\prime}$ with
\begin{align}
\hat{\Phi}^{\prime} := \frac{\hbar}{2m i \sqrt{n_0}} 
(\tilde{\phi} - \tilde{\phi}^{\dagger}).
\end{align}
Note that the field $\hat{\Phi}^{\prime}$ is the velocity potential 
in the first order approximation with respect to fluctuation fields.

Taking the difference of eqs. (\ref{eq:B-dG-tilde-0}) and (\ref{eq:B-dG-tilde}), 
we see that the operators $\hat{\rho}^\prime$ and $\hat{\bf j}^\prime$ obey the conservation law
\begin{align}
\partial_t \hat{\rho}^\prime + \nabla\cdot \hat{\bf j}^\prime = 0. \label{eq:conserv-1}
\end{align}
Therefore, the continuity equation is satisfied within first order terms for fluctuation fields.
Under the hydrodynamic approximation, in which the density gradient 
is smooth over the local healing length $\xi= \hbar / \sqrt{2m U_0 n_0}$, 
the sum of eqs. (\ref{eq:B-dG-tilde-0}) and (\ref{eq:B-dG-tilde}) yields
\begin{align}
\hat{\rho}^{\prime} \simeq - \frac{m}{U_0} 
( \partial_t + {\bf v}_0 \cdot \nabla) \hat{\Phi}^\prime.
\end{align}
Then, eq. (\ref{eq:conserv-1}) can be rewritten as
\begin{align}
\left( \partial_t + \nabla \cdot {\bf v}_0 \right) 
\frac{m}{U_0}(\partial_t + {\bf v_0} \cdot \nabla ) \hat{\Phi}^{\prime} 
- \nabla \cdot \left(\frac{m}{U_0}c_s^2 \nabla \hat{\Phi}^{\prime}\right) 
\simeq 0, \label{eq:rela-org}
\end{align}
where $c_s := \sqrt{U_0 n_0 /m}$.
Now, we introduce a symmetric tensor
\begin{align}
g_{\mu\nu} = \frac{\Lambda m c_s}{U_0} \left(
\begin{array}{cc}
- ( c_s^2 - v_0^2) & - v_{0a} \\
- v_{0b} & \delta_{ab}
\end{array}
\right), \label{eq:effectivemetric-Unruh}
\end{align}
where  $\Lambda$ is a constant and $a,b =1,2,3$. 
We can then write eq. (\ref{eq:rela-org}) as
\begin{align}
\frac{1}{\sqrt{-{\rm det}g}} \partial_{\mu} 
\left[ \sqrt{-{\rm det}g} g^{\mu\nu} \partial_{\nu} \hat{\Phi}^\prime \right] = 0.
\end{align}
If we identify $g_{\mu\nu}$ with a spacetime metric through 
$ds^2 = g_{\mu\nu} dx^{\mu} dx^{\nu}$, 
the field $\hat{\Phi}^\prime$ obeys the equation of motion for 
a minimally coupled scalar field in curved spacetime with 
the metric (\ref{eq:effectivemetric-Unruh}), and can be seen to be a field in curved spacetime.

The equation for the field $\hat{\Phi}^{\prime}$ given in (\ref{eq:rela-org}) 
was already derived as a perturbation equation of the GP equation, 
see for example \cite{Jain07,Weinfurtner:2007dq}.
In this section, we have rederived it directly from time dependent BdG 
equations as an extension of analysis given in \cite{Leonhardt02-1}.
This derivation is physically equivalent to the original derivation of 
analogue metric given in \cite{Garay-PRL}.

\section{Construction of quantum field theory in curved spacetime}
\label{sec:QFT-BdG}

Now, the field $\hat{\Phi}^{\prime}$ is found to be 
a corresponding field in analogue spacetime.
In this section, however, we show that the number of quanta described by 
$\hat{\Phi}^{\prime}$ in the analogue spacetime does 
not quantitatively correspond to that of quasiparticles in BEC, 
and there is an exact expression for a quantum field in analogue spacetime 
which is different from $\hat{\Phi}^{\prime}$.
To address this point, 
we construct quantum field theory on the analogue spacetime by using the BdG theory.

For that purpose, we introduce linear transformation 
and separate operators and wave functions so that the operators $\alpha_j$ 
and  $\alpha_j^{\dagger}$ respectively become annihilation and creation operators, as
\begin{align}
\tilde{\phi} = \sum_j [ u_j \alpha_j -v_j^\ast \alpha^{\dagger}_j ], 
\quad \tilde{\phi}^{\dagger} = \sum_j [ u_j^\ast \alpha_j^{\dagger} -v_j \alpha_j ].
\label{expansion-operators}
\end{align}
The operators $\alpha_j$ and  $\alpha_j^{\dagger}$ are set to satisfy
 the boson commutation relation $[\alpha_j, \alpha_k^{\dagger}] = \delta_{jk}$.
The wave functions $u_j$ and $v_j$ obey the following matrix equations:
\begin{align}
i \hbar \partial_t \left(
\begin{array}{c}
u_j \\ v_j
\end{array}
\right) = \left(
\begin{array}{cc}
\tilde{W} & - U_0 n_0  \\ U_0 n_0  & - \tilde{W}^\ast
\end{array}
\right) \left(
\begin{array}{c}
u_j \\ v_j
\end{array}
\right),  \label{eq:u_v}
\end{align}
where $\tilde{W} := \tilde{L} + 2 U_0 n_0 + \hbar \partial_t S $.
They then satisfy the orthonormal conditions
\begin{align}
\begin{array}{c}
\displaystyle
\int d^3x \: (u_j^\ast u_k -v_j^\ast v_k) = \delta_{jk}, \\ \displaystyle
\int d^3x \:(u_j u_k -v_j v_k) = 0.
\end{array}
\label{eq:orthonormal-BdG}
\end{align}
The operators $\alpha_j$ and $\alpha_j^{\dagger}$ describe 
the annihilation and creation of Bogoliubov quasiparticles, respectively.
Here, we identify these with annihilation and creation operators of quanta 
in the analogue spacetime.

The field $\hat{\Phi}^\prime$ is expanded in terms of operators by use 
of the expansion (\ref{expansion-operators}). 
\begin{align}
\hat{\Phi}^\prime=\sum_j \left[ \alpha_j \tilde{f}_j + \alpha^{\dagger}_j \tilde{f}^{\ast}_j \right], 
\label{eq:expand-Phi}
\end{align}
where $\tilde{f}_j$ is defined as
\begin{align}
\tilde{f}_j := \frac{\hbar}{2 mi \sqrt{ n_0}} (u_j + v_j). \label{eq:def-tilde-f}
\end{align}
Firstly, we check that the coefficient functions $\tilde{f}_j$ 
satisfies the relativistic wave equation with the metric (\ref{eq:effectivemetric-Unruh}). 
Eq. (\ref{eq:GP-S}) leads to
\begin{align}
\tilde{W} = K^\prime + U_0 n_0 -i\hbar D_v,
\end{align}
with these definitions $K^\prime := K - n_0^{-1/2} ( K n_0^{1/2} )$ and 
\begin{align}
D_v := {\bf v}_0 \cdot \nabla + \frac{1}{2}(\nabla \cdot {\bf v}_0).
\end{align}
The matrix equation (\ref{eq:u_v}) then gives equations for 
the functions $u_j + v_j$ and $u_j-v_j$ as 
\begin{align}
i \hbar \left(\partial_t + D_v  \right) ( u_j + v_j ) 
& \approx 2 U_0 n_0 (u_j - v_j), \label{eq:u+v} \\
i \hbar \left( \partial_t + D_v \right) ( u_j - v_j ) 
& = K^\prime (u_j + v_j), \label{eq:u-v}
\end{align}
where the hydrodynamic approximation,
\begin{align}
K^\prime + 2 U_0 n_0 \approx  2 U_0 n_0,
\end{align}
has been used.
From eq.(\ref{eq:u+v}),
\begin{align}
u_j - v_j = \frac{i \hbar}{2 U_0 n_0} \left( \partial_t + D_v \right) (u_j + v_j).
\end{align}
Substituting this into eq.(\ref{eq:u-v}), we obtain
\begin{align}
(\partial_t + D_v) \frac{1}{U_0n_0} 
(\partial_t + D_v)(u_j + v_j) + \frac{2 }{\hbar^2} K^\prime (u_j + v_j) = 0.
\end{align}
The conservation of the condensate current (\ref{eq:conserv-0}) leads to 
\begin{align}
& \left( \partial_t + {\bf v}_0 \cdot \nabla \right) \frac{X}{\sqrt{n_0}} =
 \frac{1}{\sqrt{n_0}} \left( \partial_t + D_v  \right) X, \label{eq:byproduct-conserv} \\
& (\partial_t + D_v) \frac{1}{\sqrt{n_0}} X = 
\frac{1}{\sqrt{n_0}} \left(\partial_t + \nabla \cdot {\bf v}_0 \right) X, 
\label{eq:byproduct-conserv-2}
\end{align}
for any function $X$. Therefore, we obtain
\begin{align}
& \left( \partial_t + \nabla\cdot {\bf v}_0 \right) \frac{1}{U_0}
(\partial_t + {\bf v}_0 \cdot \nabla ) \tilde{f}_j  \nonumber  \\
& + \frac{2 }{\hbar^2} \sqrt{n_0} K^\prime \sqrt{n_0} \tilde{f}_j  =0.
\end{align}
Furthermore, by noting that
\begin{align}
K^\prime \sqrt{n_0} X = - \frac{\hbar^2}{2m} \frac{1}{\sqrt{n_0}} \nabla \cdot (n_0 \nabla X),
\end{align}
one finds that the functions $\{\tilde{f}_j\}$ satisfy 
\begin{align}
\partial_{\mu} \left[ \sqrt{-{\rm det} g} g^{\mu\nu} \partial_{\nu} \tilde{f}_j \right] = 0, 
\label{eq:KG-f}
\end{align}
with metric (\ref{eq:effectivemetric-Unruh}), as expected.

Now, we calculate the Klein-Gordon (KG) inner product between 
functions $\{\tilde{f}_j\}$
in the analogue spacetime, 
which is a conserved product for solutions to the wave equation in curved spacetime.
The KG product for functions $F$ and $G$ is defined as
\begin{align}
(F,G)_{KG} = -i \int_{\Sigma} (F n^{\mu} \partial_{\mu} G^\ast - G^\ast 
 n^{\mu} \partial_{\mu} F) \sqrt{{\rm det}h} \: d^3x,
\end{align}
where $n = n^{\mu} \partial_{\mu}$ is a unit normal vector 
orthogonal to the spacelike surface $\Sigma$, that is, $n$ is a future-directed timelike vector.
$\sqrt{{\rm det}h} \: d^3 x$ is an invariant volume element of the surface $\Sigma$.
$h_{ab}$ is the induced metric on $\Sigma$.
If $F$ and $G$ are solutions of the Klein-Gordon equation (\ref{eq:KG-f}), 
the value $(F,G)_{KG}$ does not depend on the choice of $\Sigma$ 
as long as $\Sigma$ is a Cauchy surface \cite{BD}.
If we choose a $t$-constant surface as $\Sigma$, the induced metric is
\begin{align}
ds^2_{\Sigma} = h_{ab} dx^a dx^b = \frac{m \Lambda}{U_0} c_s \delta_{ab} dx^a dx^b.
\end{align}
Then, we obtain $\sqrt{{\rm det}h} = (m \Lambda c_s)^{3/2}U_0^{-3/2}$.
The normal vector to the surface is
\begin{align}
n^{\mu} \partial_{\mu} = c_s^{-3/2} 
\sqrt{\frac{U_0}{m\Lambda }}
( \partial_t + {\bf v}_0 \cdot \nabla ).
\end{align}
By using eqs. (\ref{eq:byproduct-conserv}) and (\ref{eq:u+v}), 
the KG product between the coefficient functions can be calculated as
\begin{align}
(\tilde{f}_j, \tilde{f}_k)_{KG} = \frac{\Lambda \hbar }{m}
\int_{\Sigma} d^3x \: \left[ u_j u_k^\ast - v_j v_k^\ast \right]. 
\label{eq:f-uv}
\end{align}
Using the orthonormal relation (\ref{eq:orthonormal-BdG}), we obtain 
\begin{align}
(\tilde{f}_j, \tilde{f}_k)_{KG} = \frac{\Lambda \hbar }{m} \delta_{jk}.
\end{align}
Similar calculations show that 
$(\tilde{f}_j^\ast, \tilde{f}_k^\ast)_{KG} = - \frac{\Lambda \hbar }{m} \delta_{jk}$ 
and $(\tilde{f}_j, \tilde{f}_k^\ast)_{KG} = 0$.
Therefore, it is seen that the coefficient functions are normal each other.
At the same time, it is also seen that they are not orthonormal.
We find that a set of functions 
\begin{align}
\frac{\sqrt{m}}{\sqrt{\Lambda \hbar }}\tilde{f}_j,
\end{align}
is a set of orthonormal mode functions for the field $\hat{\Phi}^\prime$.
Then the number operator of quanta described by $\hat{\Phi}^\prime$ is
\begin{align}
\tilde{N}_j=\frac{\Lambda \hbar }{m} \alpha_j^{\dagger}\alpha_j.
\end{align}
The number of quanta in analogue spacetime does not correspond to that 
of Bogoliubov quasiparticles.
In order to identify these numbers, the field has to be normalized 
so as to make the coefficient functions orthonormal.
It is easily seen that a set of functions
\begin{align}
f_j := \frac{\sqrt{\hbar }}{2 i \sqrt{\Lambda m n_0}} (u_j + v_j), \label{eq:def-f}
\end{align}
is a complete set in quantum field theory in analogue curved spacetime, 
that is, these functions satisfy
\begin{align}
(f_j, f_k)_{KG} = \delta_{jk}.
\label{eq:complete-set}
\end{align}
Then the  normalized field is 
\begin{align}
\varphi = \sum_j \left[ \alpha_j f_j + \alpha^{\dagger}_j f^{\ast}_j \right] 
= \frac{\sqrt{m}}{\sqrt{\Lambda \hbar}} \hat{\Phi}^\prime. 
\label{eq:def-varphi}
\end{align}
Then, the number operator of the field $\varphi$ is 
\begin{align}
N_j=\alpha_j^{\dagger}\alpha_j.
\end{align}
We have obtained orthonormal mode functions (\ref{eq:def-f})
and normalized scalar field (\ref{eq:def-varphi}) 
in the analogue spacetime from BdG wave functions.
Equivalently, we have exactly related one Bogoliubov quasiparticle with one quantum in 
the analogue spacetime.
Now, with this normalized field $\varphi$, 
quantum field theory in curved spacetime can be applicable to
Bogoliubov quasiparticles quantitatively.

\section{Particle creation}
\label{sec:particle-creation}

In this section, we discuss a simple example of particle creation 
and derive a formula connecting particle creation spectrum and BdG wave functions.

Let us consider a condensate which is static and having no excitation at 
initial time $t = t_1$.
For $t_1 < t < t_2$, the condensate evolves dynamically and, at $t=t_2$, 
it becomes static again.
Then, the corresponding effective spacetime evolves dynamically and, in general, 
quantum field theory in curved spacetime will predict particle creation.
Therefore, the analogy implies that, Bogoliubov quasiparticles will be created at $t=t_2$.

We choose a complete set at time $t = t\sub{\lambda}$ ($\lambda=1,2$), 
$\{f_j^{(\lambda)}\}$, 
as each $f_j^{(\lambda)}$ consists of energy eigenfunctions of the BdG equations 
$u_j^{(\lambda)}$ 
and $v_j^{(\lambda)}$ at each time $t=t\sub{\lambda}$.
In this way, we define one particle state of the quantum 
by using annihilation and creation operators $\alpha_j^{(\lambda)}$ 
and $\alpha_j^{(\lambda)\dagger}$ which diagonalize the BdG Hamiltonian.

At the initial and final times $t = t\sub{\lambda}$ ($\lambda = 1, 2$), 
we expand the field using a complete set at each time as
\begin{align}
\varphi = \sum_j \left[ \alpha_j^{(\lambda)} f^{(\lambda)}_j 
+ \alpha^{(\lambda)\dagger}_j f^{(\lambda)\ast}_j \right],
\end{align}
These two sets are related to each other by linear transformations
\begin{align}
f^{(2)}_k = \sum_j\left[ A_{kj} f^{(1)}_j + B_{kj} f^{(1)\ast}_j \right].
\end{align}
Therefore, the creation and annihilation operators are related as
\begin{align}
\alpha_k^{(2)} = \sum_{j} 
\left( A_{kj}^\ast \alpha_j^{(1)} - B_{kj}^\ast \alpha_j^{(1) \dagger} \right).
\end{align}
Since the initial condensate has no excitations, 
the state for $\varphi$ is in the initial vacuum denoted by $\ket{0}_{(1)}$, 
i.e., $\alpha_j^{(1)} \ket{0}_{(1)} = 0$.
We consider the Heisenberg picture and the state for $\varphi$ is not time-dependent.
Then, the expectation value of the final number operator in 
the $j$-th mode $\hat{N}_j^{(2)} := \alpha_j^{(2)\dagger}\alpha_j^{(2)}$ is calculated as
\begin{align}
{}_{(1)} \bra{0} \hat{N}_j^{(2)} \ket{0}_{(1)} = \sum_k |B_{jk}|^2. \label{eq:formula-spectrum}
\end{align}
If $B_{kj}$ is not zero for any $k$, then particles will be created.
The expected total number of particles is thus tr($B^{\ast}B$).

We should note that solutions to the BdG equations $u_j$, $v_j$ solve 
relativistic wave equations (\ref{eq:KG-f}) through the relation (\ref{eq:def-f}).
Therefore, if the time-dependent BdG equations are solved, 
then the functions $\{f_j^{(1)}\}$ satisfy the equation of motion 
in time-dependent effective spacetime.
Therefore, time evolution of the initial complete set in analogue spacetime is solved automatically.
Then, the coefficient $B_{jk}$ is calculated as the KG product at the final time $t = t_2$,
\begin{align}
B_{jk} &= - (f_j^{(2)}, f_k^{(1)\ast})_{KG} \nonumber  \\
& = \int_{t=t_2} d^3x (u_j^{(2)} v_k^{(1)}- v_j^{(2)} u_k^{(1)}). \label{eq:formula-B}
\end{align}
In the derivation of this formula, we have used 
(\ref{eq:byproduct-conserv}) and (\ref{eq:u+v}) as in the calculation of (\ref{eq:complete-set}).
Here, it should be noted that, in the calculation of $B_{jk}$, 
not the set of functions $\{\tilde{f}_j\}$ but $\{f_j\}$ has to be considered
so that the Bogoliubov coefficient $B_{jk}$ is relevant as 
a particle creation spectrum quantitatively.
This expression for the Bogoliubov coefficient (\ref{eq:formula-B})
was obtained in \cite{Jain07} for particle creation in homogeneous BEC.
Now, the formula (\ref{eq:formula-B}) has been shown to be valid for particle 
creation in dynamical condensates even when it is inhomogeneous,
because, in our formulation, any simplification for the Bogoliubov mode functions 
is not needed to be assumed. 
Therefore, our formulation is much more powerful.
It is powerful enough to show that the formula (\ref{eq:formula-B}) is 
valid for general cases. 
In this way, the spectrum of particle creation 
(\ref{eq:formula-spectrum}) can be expressed in terms of the BdG wave functions 
via the formula (\ref{eq:formula-B}).

Finally, we comment that the total number of particles
\begin{align}
N = \int d^3x \: \bracket{\psi^\dagger \psi} = \int d^3x \: 
(|\Psi|^2 + \bracket{\phi^\dagger \phi}),
\end{align}
is not conserved. $\partial_t N \neq 0$ in our model, 
since the conservation is closed only in the GP equation for the condensate particles;
\begin{align}
\partial_t \int d^3x \: |\Psi|^2 = 0.
\end{align}
To overcome this unphysical result, we have to account for 
the quantum backreaction which is the effect of the dynamical development of 
the Bogoliubov field to the BEC.
In this paper, 
we assume that the quantum backreaction 
can be neglected when the number of the BEC particles far exceeds 
the number of Bogoliubov quasiparticles.

\section{Numerical simulation}
\label{sec:numerical}

Now, we show a numerical simulation giving particle creation 
in an expanding and shrinking BEC in the trapping potential 
$V\sub{ext}({\bf x}) = m (\omega_x^2 x^2 + \omega_y^2 y^2 + \omega_z^2 z^2) / 2$, 
where $\omega_x$, $\omega_y$, and $\omega_z$ are the trapping frequencies of the potential 
along the $x$, $y$, and $z$-axes respectively.
We consider the case in which a BEC is tightly trapped only in the $x$-axis; 
$\omega_x \gg \omega_y, \omega_z$, and apply the one-dimensional simulation 
by neglecting the spatial derivatives of the $y$ and $z$-axes.
We find the stationary solution of the GP equation at the trapping frequency 
$\omega_x = \omega\sub{i}$, and then change the frequency to $\omega_x = \omega\sub{f}$ 
at $t = t_1 = 0$ to make the expanding and shrinking BEC.
This situation can be interpreted as a cosmological expansion and contraction, 
because, in the expansion of condensate for example, 
the sound velocity becomes small and the time interval $\Delta t$ 
during which a particle on the condensate can travel over it,
\begin{align}
\Delta t := 2\int^{R_{TF}}_0 c_s^{-1} dx,
\end{align}
becomes long, where $R_{TF}$ is the Thomas-Fermi edge of the condensate. 
This implies that the size of the analogue spacetime measured by the sound velocity also expands.
Furthermore, the sound velocity is not constant with respect to the spatial coordinate, 
which implies that the analogue spacetime is inhomogeneous.
Therefore, the analogue spacetime corresponds to an inhomogeneous expanding universe. 
Similarly, a shrinking BEC corresponds to an inhomogeneous contracting universe.

Although the condensate starts to move due to changing the trapping frequency, we are not interested in particle creation caused by such an external manipulation but in spontaneous one as a quantum field theoretical phenomenon.
For that purpose, we set vacuum for Bogoliubov quasiparticles at time just 
after changing the frequency, say $t=t_1+0$.
This is the initial condition in this simulation.
After $t=t_1+0$, therefore, the system is isolated without external manipulation, and particle creation can be regarded as a consequence of time evolving analogue spacetime.

The spectrum of particle creation $\sum_k |B_{jk}|^2$ can be calculated as follows. 
Firstly, we calculate the time evolution of $u_j^{(1)}$ and $v_j^{(1)}$ with 
that of the condensate wave function $\Psi$ by using eq. (\ref{eq:u_v}), 
starting from the quasi-steady solution at $t = t_1+0$:
\begin{align}
E_j^{(\lambda)} \left(
\begin{array}{c}
u_j^{(\lambda)} \\
v_j^{(\lambda)}
\end{array}
\right) = \left(
\begin{array}{cc}
\tilde{W} & - U_0 n_0 \\
U_0 n_0 & - \tilde{W}^\ast
\end{array}
\right) \left(
\begin{array}{c}
u_j^{(\lambda)} \\
v_j^{(\lambda)}
\end{array}
\right). \label{eq:stationary-u_v}
\end{align}
Then, at $t = t_2$, we obtain $u_j^{(1)}$ and $v_j^{(1)}$ 
after the time evolution and $u_j^{(2)}$ and $v_j^{(2)}$ 
by using eq. (\ref{eq:stationary-u_v}) again.
Finally, we obtain the spectrum of particle creation from eq. (\ref{eq:formula-B}).

As the physical parameters, we use $m = 1.44 \times 10^{-25}$ kg, 
$\omega\sub{i} = 150 \times 2 \pi$ Hz, $\omega\sub{f} = \omega\sub{i}/\sqrt{2}$, 
and $\omega_y = \omega_z = \omega_\perp = 50 \times 2 \pi$ Hz 
considering a BEC experiment of $^{87}$Rb atoms, the total number of which 
is set to be $N = 2.5 \times 10^{6}$.
As the coupling constant of atomic interactions $U_0$, we use $s$-wave scattering:
\begin{align}
U_0 = 4 \left[ \frac{\hbar^{3} a^{3/2}
\omega_\perp^3 N^{3/2}} {15 m^{1/2} \omega_x} \right]^{2/5},
\end{align}
with a scattering length of $a = 5.61$ nm.
As a numerical method, we use the fully dealiased Chebyshev-Galerkin method 
in space $-16 a_x \le x \le 16 a_x$ with the Dirichlet boundary condition 
and the Runge-Kutta-Gill method in time with $\Delta t = 1 \times 10^{-4} / \omega\sub{i}$, 
where $a_x = \sqrt{\hbar / m \omega\sub{i}}$ is the characteristic length defined 
by the trapping potential.

\begin{figure}[tbh]
\centering
\begin{minipage}{0.49\linewidth}
\centering
\includegraphics[width=0.99\linewidth]{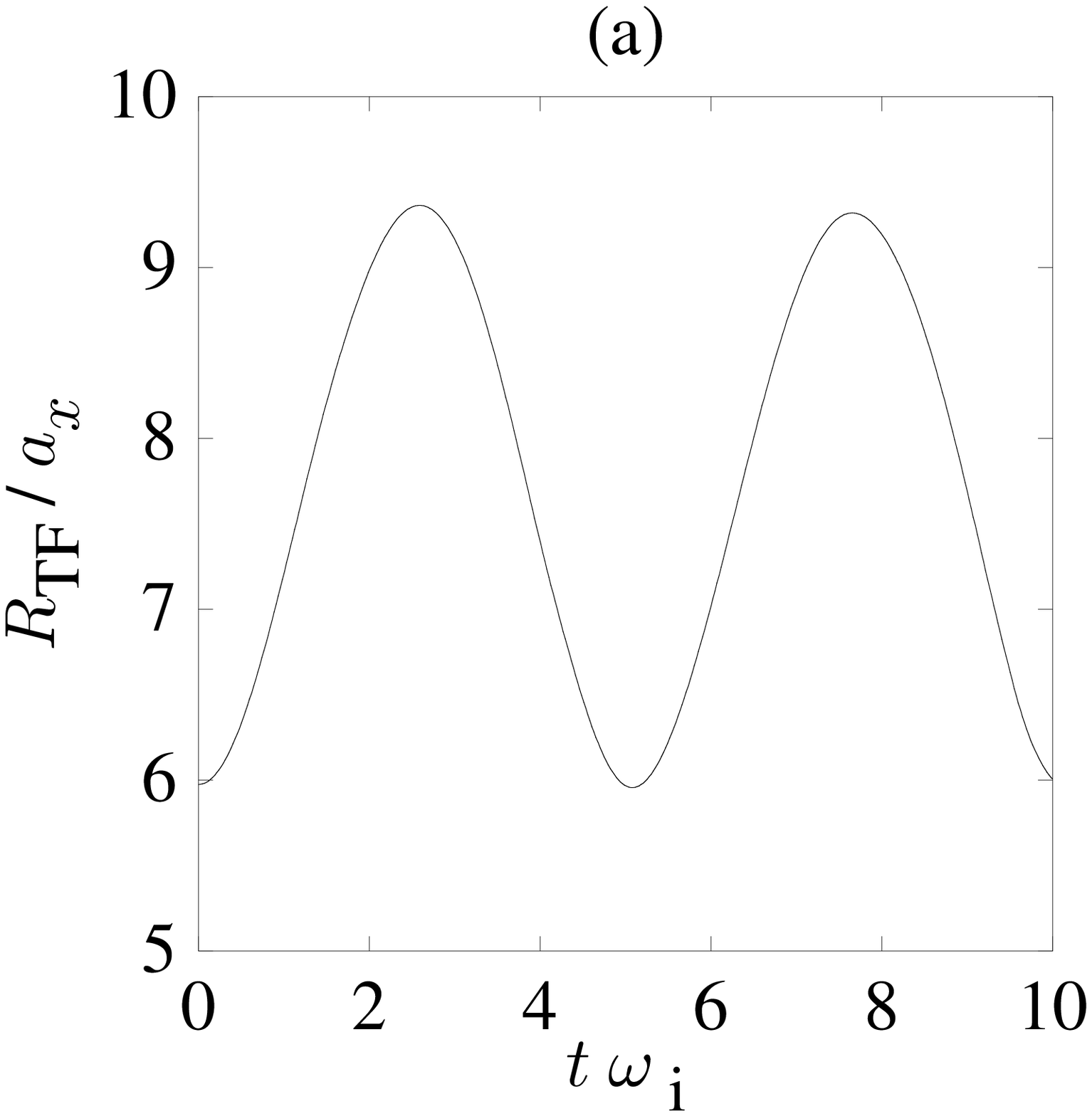}
\end{minipage}
\begin{minipage}{0.49\linewidth}
\centering
\includegraphics[width=0.99\linewidth]{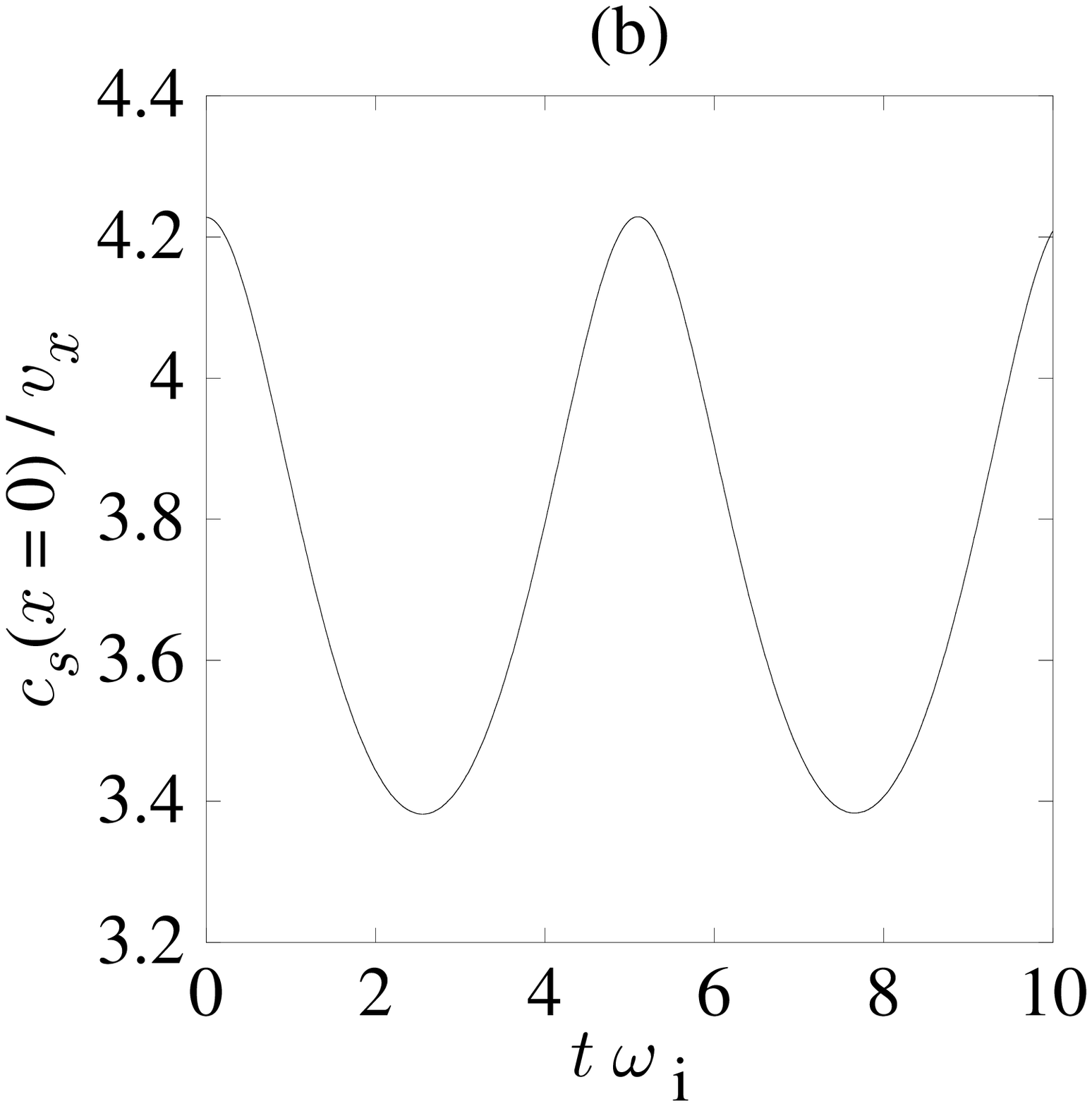}
\end{minipage}
\caption{\label{fig:TFedge} Dependence of the Thomas-Fermi edge $R\sub{TF}$ 
(a) and the sound velocity $c_s = \sqrt{U_0 n_0(x =0)/m}$ (b) at the center on time. 
Here, $a_x = \sqrt{\hbar / m \omega\sub{i}}$, $v_x = \sqrt{\hbar \omega\sub{i} / m}$ 
are the characteristic length and velocity defined by the trapping potential.}
\end{figure}

Figure \ref{fig:TFedge}(a) shows the time development of 
the Thomas-Fermi edge $R\sub{TF}$ of the condensate fitted by a harmonic function.
Dynamics of the condensate is almost periodic, and its period is 
$T \simeq 5.10 / \omega\sub{i}$ which is consistent 
with that of the breathing mode $ \sim 2 \pi / (\sqrt{3/2} \omega\sub{i})$.
The sound velocity $c_s$ decreases with an increase in 
$R\sub{TF}$ as shown in Fig. \ref{fig:TFedge}(b), also becoming periodic in time.

\begin{figure}[tbh]
\centering
\begin{minipage}{0.49\linewidth}
\centering
\includegraphics[width=0.99\linewidth]{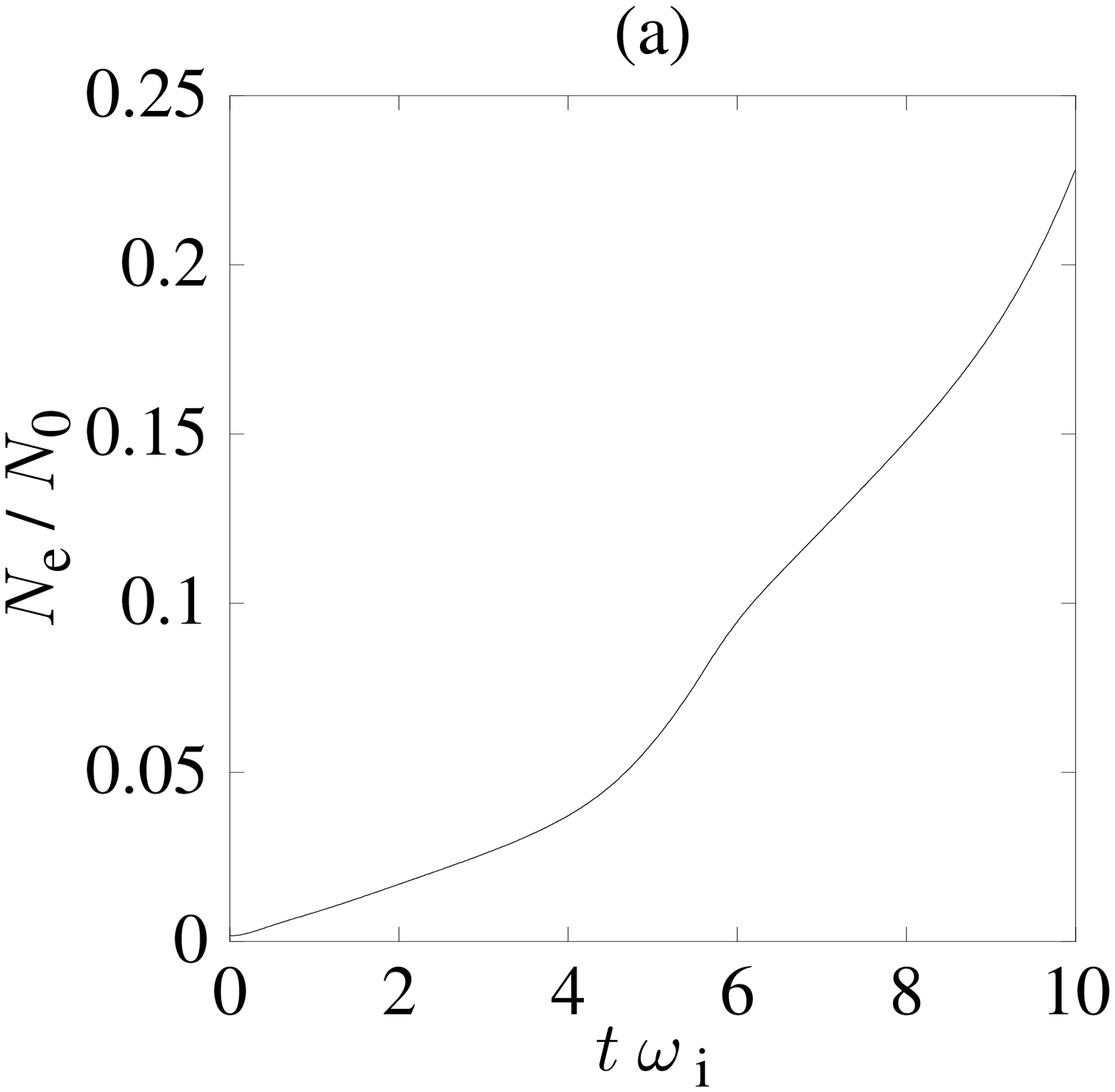}
\end{minipage}
\begin{minipage}{0.49\linewidth}
\centering
\includegraphics[width=0.99\linewidth]{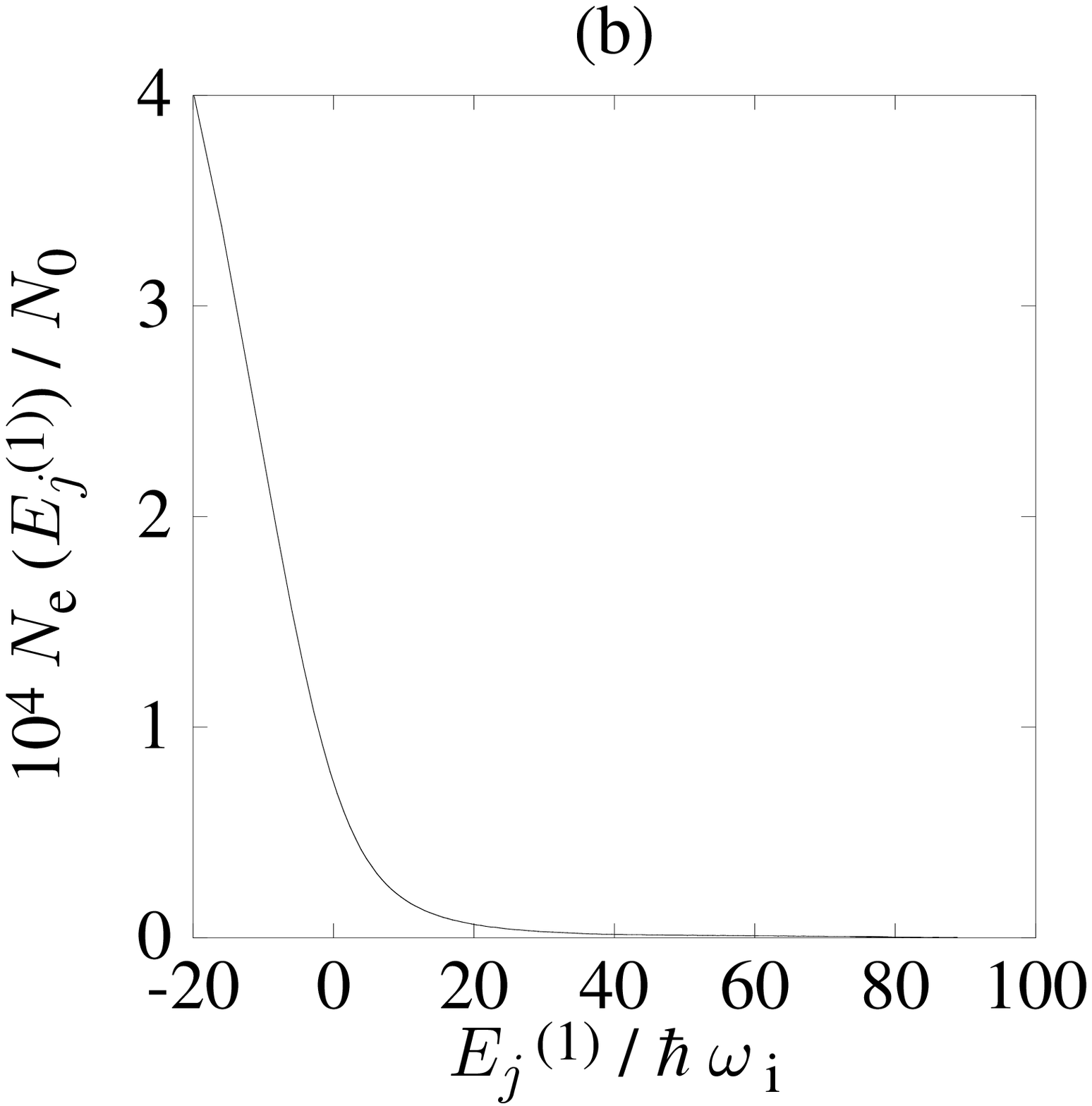}
\end{minipage}
\caption{\label{fig-noncondensate} Dependence of the number of 
the Bogoliubov quasiparticles on time (a) and its spectrum (b).}
\end{figure}

The decrease in the sound velocity shown in Fig. \ref{fig:TFedge}(b) 
corresponds to expansion of the analogue spacetime.
We select the time at which the size of condensate becomes maximum as the final time $t_2$, 
i.e., $t_2:=T / 2 \simeq 2.55 / \omega\sub{i}$, 
and calculate $u_j^{(\lambda)}$ and $v_j^{(\lambda)}$ ($\lambda=1,2$).
Before calculating the spectrum of the particle creation, 
we need to verify whether the effect of quantum backreaction can be neglected, as discussed above.
Figures \ref{fig-noncondensate}(a) and (b) show 
the time dependence of the total number of non-condensate particles; 
\begin{align}
N\sub{e} & = \int dx \: \bracket{\phi^\dagger \phi}_{(1)} \nonumber \\
& = \int dx \: \sum_j N\sub{e}(E_j^{(1)}) = \int dx \: \sum_j |v_j^{(1)}|^2,
\end{align}
and its spectrum at $t = t_2$.
Both figures reveal that the number of Bogoliubov quasiparticles is 
small enough relative the number of BEC $N_0 :=\int dx \: |\Psi|^2$ 
that the quantum backreaction will not alter the result qualitatively.

\begin{figure}[tbh]
\centering
\begin{minipage}{0.49\linewidth}
\centering
\includegraphics[width=0.99\linewidth]{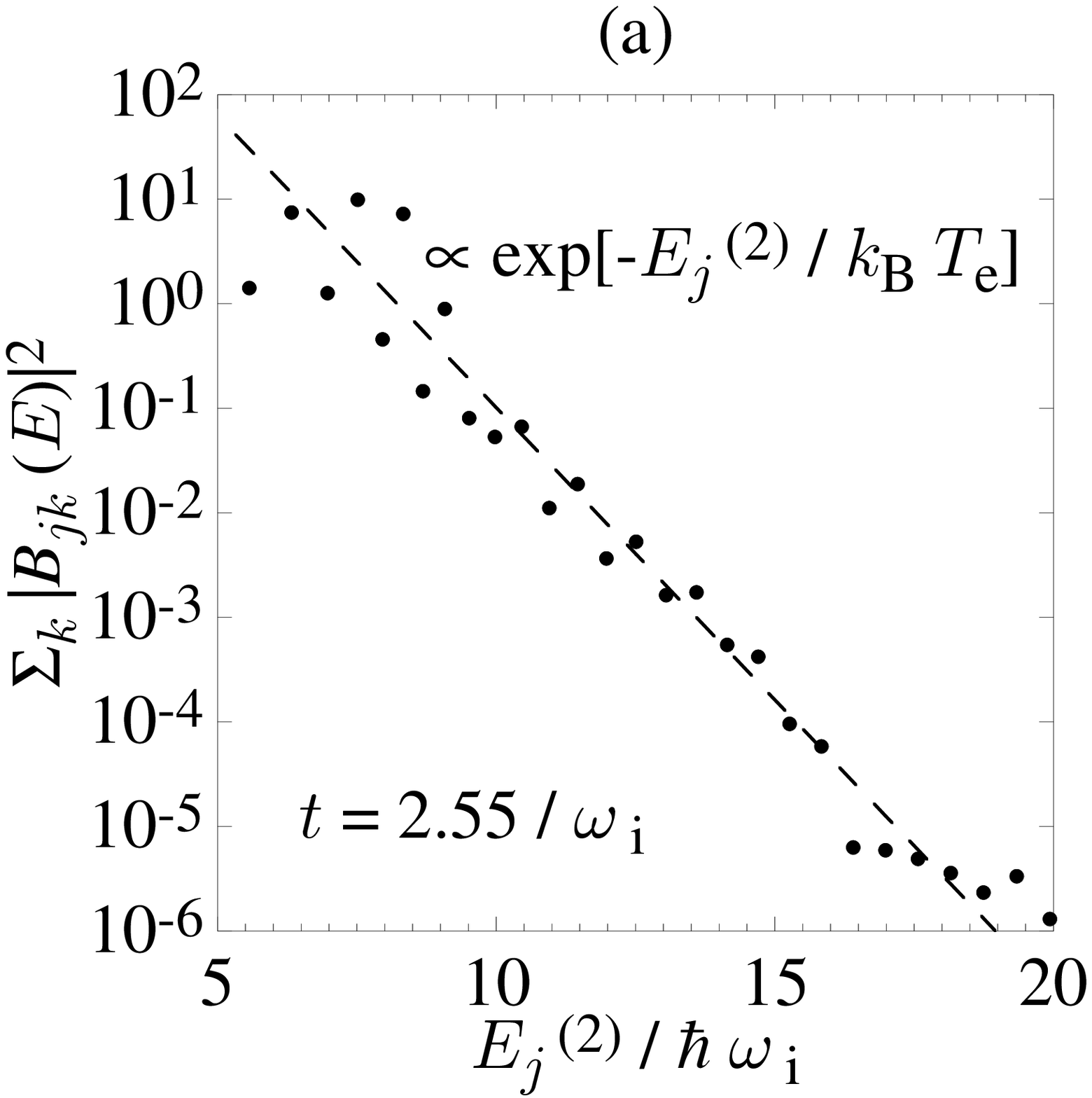}
\end{minipage}
\begin{minipage}{0.49\linewidth}
\centering
\includegraphics[width=0.99\linewidth]{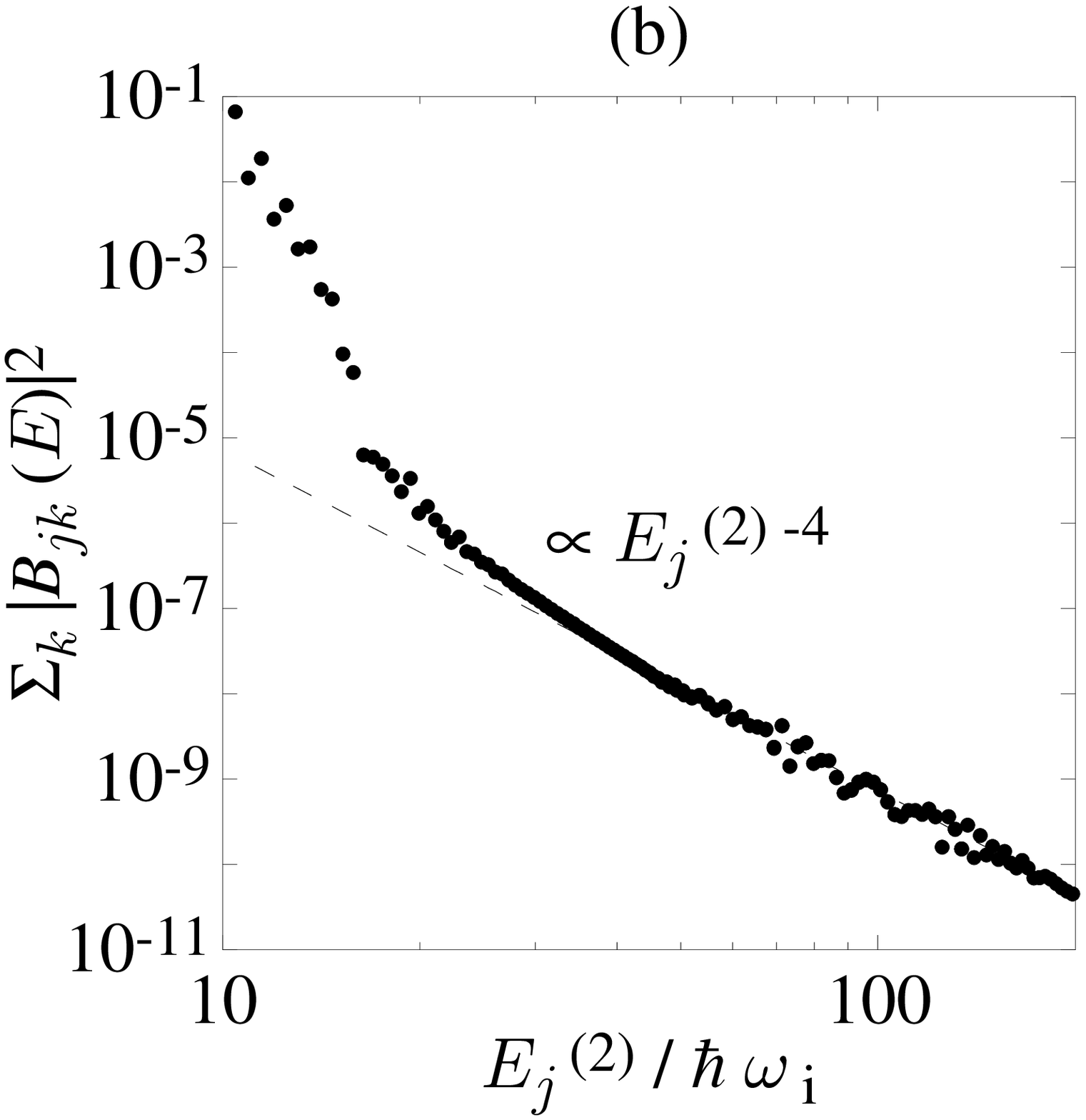}
\end{minipage}
\caption{\label{fig:Bnm} The spectrum of particle creation at 
$t = 2.55 / \omega\sub{i}$ in the low energy region 
$5 \hbar \omega_{i} < E_j^{(2)} < 20 \hbar \omega_{i}$ (a) 
and $10 \hbar \omega_{i} < E_j^{(2)} < 200 \hbar \omega_{i}$ (b).}
\end{figure}

Next, we investigate its particle creation. Figure \ref{fig:Bnm} shows 
the spectrum of the particle creation $\sum_k |B_{jk}|^2$.
In the low energy region $E_j^{(2)} < \mu$ corresponding to Fig. \ref{fig:Bnm}(a), 
the spectrum has the thermal Maxwell-Boltzmann distribution: 
$\sum_k|B_{jk}|^2 \propto \exp(-E_j^{(2)} / k\sub{B} T\sub{e})$ with the temperature 
$k_{B} T\sub{e} \simeq 0.777 \hbar \omega\sub{i}$.
This result means that the spectrum of the particle creation in 
the analogue expanding universe obeys the thermal distribution with some temperature $T\sub{e}$.
It is widely believed that the thermal spectrum of particle creation
is caused by formation of a static horizon. However, at the time $t_2$, there is no sonic horizon.
This particle creation is quite nontrivial.

To confirm this experimentally, a BEC is needed at a temperature lower than $T\sub{e}$ , to ensure that it is not be submerged by thermal noise.
Using realistic parameters, $T\sub{e}$ is estimated to be 5.60 nK 
which has been already achieved in many BEC experiments. 
On the other hand, in the high energy region $E_j^{(2)} \gg \mu$ shown in 
Fig. \ref{fig:Bnm} (b), the spectrum exhibits power-law behavior 
$\sum_k |B_{jk}|^2 \propto E_j^{(2) -4}$.
This power-law behavior, however, seems to be difficult to detect 
because particle creation in this region is very weak.
In the high energy region, the modes are not hydrodynamic and the 
analogy breaks down. The power-law behavior might be explained by 
trans-Planckian physics in analogy\cite{Corley:1996ar,Weinfurtner:2007dq}.

Note that the particle creation shown in this simulation occurs 
spontaneously in the sense that the initial state for the excitation field 
has no excitation at the time just after the changing of the 
trapping frequency, and there is no external manipulation after the condensate
starts to move. 
This is contrasted to the previous simulation giving homogeneous cosmological
analogue \cite{Jain07}, 
in which the external manipulation like changing strength of the atomic interaction
is required.

\section{Summary}
\label{sec:summary}

In conclusion, we have formulated quantum field theory in analogue spacetime 
based on the BdG equations.
In this analogy, quanta in curved spacetimes are explicitly exactly related with 
Bogoliubov quasiparticles on condensates. 
It has been demonstrated that, the orthonormal relations for Bogoliubov modes 
correspond to that for quanta in effective spacetime with respect to the KG product.
We derive a simple formula for the particle creation spectrum in terms of BdG wave 
functions, which can be applied to simple dynamical evolution whose initial 
and final condensations are quasi-static.
Furthermore, we calculate the particle creation in the analogue expanding universe 
by numerically solving time-dependent BdG equations for an expanding BEC.
The spectrum obtained is consistent with the thermal Maxwell-Boltzmann distribution for 
the temperature $T\sub{e} \simeq 5.60$ nK, which is experimentally accessible. 
This supports the experimental testing of particle creation in an expanding universe 
by using a BEC.

Because we neglected the effect of the quantum backreaction from the Bogoliubov field 
to the BEC, we are unable to calculate the precise amount of particle creation.
To overcome this deficiency, we are extending the
model to incorporate the quantum backreaction and will report on this in near future.
Furthermore, numerical simulations of particle creations in dynamical BEC corresponding 
to Hawking radiation will be reported in future studies.

{\it Acknowledgments:} 
MK acknowledges JSPS Research Fellowships for Young Scientists (Grant No. 207229). 
MT is supported in part by a Grant-in-Aid for Scientific Research from JSPS (Grant No. 18340109) and
by a Grant-in-Aid for Scientific Research on Priority Areas from MEXT (Grant No. 17071008).
HI is supported by a Grant-in-Aid for Scientific Research 
Fund of the Ministry of Education, Science and Culture of Japan (Grant No. 19540305).

\end{document}